\documentstyle[prl,aps]{revtex}

\begin{document}
\twocolumn
%\draft
\title{Laser-induced Rotation of a Trapped Bose-Einstein Condensate}
\author{Karl-Peter Marzlin and Weiping Zhang}
%\\[2mm]
\address{
School of Mathematics, Physics,
Computing and Electronics, Macquarie University, Sydney, NSW 2109,
Australia
}
\maketitle
%%%%%%%%%%%%%%%%%%%%%%%%%%%%%%%%%%%%%%%%%%%%%%%%%%%%%%%%%%%%%%%%
\begin{abstract}
In this letter, atom optic techniques are proposed to control
the excitation of a Bose-Einstein condensate in an atomic trap.
We show that by employing the dipole potential induced by four
highly detuned travelling-wave laser beams with appropriate phases
and frequencies, one can coherently excite a trapped
Bose-Einstein condensate composed of ultracold alkali
atoms into a state rotating around the trap center. The connection
to vortex states is discussed.
\end{abstract}
%%%%%%%%%%%%%%%%%%%%%%%%%%%%%%%%%%%%%%%%%%%%%%%%%%%%%%%%%%%%%%%%
$ $ \\
%\narrowtext
The experimental realizations of Bose-Einstein condensition in an atomic
trap \cite{experimente} have led to a broad interest in the properties of
coherent ultracold atomic gases. The comparison of trapped condensates
to superfluid Helium suggests the possibility to study and analyse the
rotational properties of the condensate \cite{rotation}. This
is usually made by exciting states in which the
atoms move collectively around
one or several vortex lines. Recently the properties of vortex states of
a trapped Bose-Einstein condensate has been studied theoretically
\cite{vortex}.

In this letter we propose to employ the atom optic techniques to study
the excitation of rotating states, being a coherent superposition
of vortex states, from the ground state of the trap.
In atom optics, the center-of-mass motion of atoms can be manipulated
by light-induced gradient forces. We apply the same idea to a trapped
Bose-Einstein condensate which can be described by a single macroscopic
wave function. The interaction of light waves with ultracold atomic
samples composed of Bose-Einstein condensates has well been studied
in recent years \cite{zhang1}-\cite{you}. The vector nonlinear
stochastic Schr\"odinger equations describe the dynamics of the
ground- and excited-state atomic wave packets in
the laser fields\cite{zhang2}. For large laser detuning,
adiabatic elimination of the excited-state component leads to
a nonlinear scalar Schr\"odinger equation
for the ground-state macroscopic wave function $\psi_g$
of the condensate \cite{peter},
\begin{eqnarray}
i \hbar \frac{d \psi_g(\vec{x},t) }{dt}& =& (H_{c.m.} + V_L(\vec{x}))
  \psi_g(\vec{x},t)  \nonumber \\ & & +
  \lambda |\psi_g(\vec{x},t)|^2 \psi_g(\vec{x},t)\; .
\label{gpe} \end{eqnarray}
The parameter $\lambda$ is connected to the s-wave scattering length
of the atoms in the presence of the laser field. For simplicity we
ignore the effects of the nonlinearity by setting $\lambda = 0$
in this paper. The incorporation of the nonlinear interaction in
the proposed scheme will be the subject of a further publication.
The center-of-mass Hamiltonian $H_{c.m.}$ includes the
kinetic energy and the harmonic potential of the trap.
According to the usual treatment of the quantum mechanical
harmonic oscillator the Hamiltonian can be expressed as
\begin{equation}
H_{c.m.} = \hbar \omega_\perp \{ a_x^\dagger a_x + a_y^\dagger
  a_y + 1 \} + \hbar \omega_z \{a_z^\dagger a_z +1/2 \} \; ,
\label{Heq} \end{equation}
where $a_i := [ x_i/R_i + i R_i p_i/\hbar ]/\sqrt{2}$,
$i=x,y,z$, is the annihilation operators for the harmonic
oscillator in the i-direction and $R_i := \sqrt{\hbar/(M
\omega_i)}$ is the trap size parameter.
$M$ is the mass of the atoms, $\omega_\perp$ is the trap
frequency for the x- and y-direction, and $\omega_z$ is the
trap frequency in the z-direction. Because the trap is cylindrically
symmetric around the z-axis, the Hamiltonian $H_{c.m.}$ commutes with
the orbital angular
momentum operator $J_3 = x p_y - y p_x$. This is a natural result of
the conservation of orbital angular momentum along the z-directon
in the cylindrically symmetric trap in the absence of the laser fields.
To excite the rotating states from a Bose-Einstein condensate
occupying the ground state of the trap which is the ground state
$|0 \rangle$ of the harmonic oscillators, one must introduce an
additional external force in the x-y plane to break down the
rotational symmetry of the ground state around the z-axis
and create an angular momentum for a rotation of
the ground state. The laser-induced dipole potential
$V_L(\vec{x})$ in equation (\ref{gpe}) can be used for this purpose.
Before working out the exact form of the
laser-induced dipole potential required for the excitation of
vortex states, we need to understand the properties of the
rotating states for the ground-state condensate. A vortex state
rotating around the axis of symmetry is a state in which
the macroscopic wave function
$\psi_g(\vec{x})$ has the form $\exp (i n \varphi) \chi (r,z)$,
where $\varphi$ is the angle and $r$ is the radius in the
x-y-plane and $n$ is an integer.
It therefore can be characterized
as being an eigenstate of $J_3$ with eigenvalue $n \hbar$,
and of $H_{c.m.}$ as well.
To deal with rotating states it is convenient to use the complete
set of states $|v \rangle = (c_-^\dagger)^l (c_+^\dagger)^m |0 \rangle
/\sqrt{l! m!}$ to decompose the wavefunction of the condensate.
Here $|0 \rangle $ denotes the common ground state
of the three-dimentional harmonic trap, and the {\em vortex
annihilation operators} $c_\pm$ are given by
\begin{equation}
c_\pm = \frac{1}{\sqrt{2}} \{ a_x \pm i a_y \} \; .
\label{vortexan} \end{equation}
Using their commutators with $H_{c.m.}$ and $J_3$ one can show
that $c_\pm$ both annihilate one quantum $\hbar \omega_\perp$ of
phonon energy and change the angular momentum by $\pm \hbar$.

In terms of the properties of the rotating states, we conclude that
if the laser-induced dipole potential contains the appropriate
combination of the vortex operators ($c_\pm$,$c^\dagger_\pm$),
one can create a superposition of vortex states from the
ground-state condensate.
To obtain such a dipole potential, we propose to employ four
travelling-wave laser beams with the spatial configuration as
shown in Fig.1.

%%%%%%%%%%%%%%%%%%%%%%%%%%%%%%%%%%%%%%%%%%%%%%%%%%%%%%%%%
The dipole potential induced by these laser beams has the
following form \cite{zhang2,peter}
\begin{equation}
  V_L(\vec{x}) = \frac{\hbar}{4(\bar{\Delta} +
  i \gamma/2)} \sum_{a,b=1}^4 \Omega_a(\vec{x})
  \Omega_b^*(\vec{x}) e^{-i(\omega_a -\omega_b)t}
\label{veq} \end{equation}
where we assume that all lasers are highly
detuned from the transition frequency between the internal
ground and excited state. In this case the excited state has only
a small population and can be adiabatically eliminated.
In equation (\ref{veq}) the symbols
$\Omega_a (\vec{x}) := \vec{d} \cdot
\vec{E}_a^{(+)}(\vec{x}) /\hbar \; , \; a=1,\ldots ,4$
denote the Rabi frequencies for the laser beams with frequencies
$\omega_a$. $\vec{d}$ is the matrix element of dipole moment of
the two-level atoms, and $\vec{E}_a^{(+)}(\vec{x})$ is the positive
frequency part of the $a$th laser's electric field, and the average
detuning of the
lasers is given by $\bar{\Delta}$. We assume $\omega_a -\omega_b$
to be much smaller than $\bar{\Delta}$ so that the difference
between the true detuning of a laser and $\bar{\Delta}$ is of
higher order in $1/\bar{\Delta}$. The spontaneous emission rate
$\gamma$ describes the incoherent loss of atoms from the condensate.
The loss can be neglected in the time scale considered here for the case
$\Delta \gg \gamma$.

For the configuration shown in Fig.1, the two laser beams
propagating along the x- and y-direction have the same frequency
$\omega_1= \omega_2 \equiv \omega $. Their phases are chosen to be
equal at the origin so that the Rabi frequencies
of the two lasers have the form
$\Omega_1(\vec{x}) = \Omega_0 \exp [i k x]$
and $\Omega_2 (\vec{x}) = \Omega_0 \exp [i k y] $.
The other two laser beams are chosen to have the frequency
$\omega_3 = \omega_4 \equiv \omega^\prime$ and propagate in
directions slightly different
from the x- and y- direction with a small angle $\theta$.
The frequency difference $\Delta \omega := \omega -\omega^\prime$
is assumed to be so small that the difference in the wavelength
between the two pairs of laser beams can be neglected. In
terms of the geometry shown in Fig.1,
the wavevector of the third laser beam is given by
$k \cos (\theta) \vec{e}_x - k \sin (\theta) \vec{e}_y \approx
k \vec{e}_x - \delta k \vec{e}_y$, where $\delta k := k \theta$
denotes the deviation of the wavevector.
Similiarly the 4th laser's wavevector is
given by $k \vec{e}_y - \delta k \vec{e}_x$.
The phases are chosen so that laser 4 is in phase with lasers 1 and 2
at the origin whereas the phase of laser 3 is shifted by
$-\pi/2$. This phase shift is essential for the scheme to work
since in the final Hamiltonian it will lead to a $\pi/2$
phase difference between $a_x$ and $a_y$, which exactly realizes
the vortex opertors (\ref{vortexan}).
The third Rabi frequency is then given by
$\Omega_3(\vec{x}) = \exp [-i \pi/2] \Omega_0 \exp [i (k x -\delta k y)]$
and the fourth by
$\Omega_4(\vec{x}) = \Omega_0 \exp [i (k y - \delta k x)] $.
This arrangement of laser beams produces the dipole potential
\begin{eqnarray}
V_L(\vec{x})& =& \frac{\hbar |\Omega_0|^2}{4 \bar{\Delta}} \Big \{
  2 + e^{-i k(x-y)} [1+i e^{-i \delta k (x-y)}] \nonumber \\ & &
  + e^{-i \Delta \omega t} \big [ e^{i \delta k x} + i e^{i\delta
  k y} +i e^{-i k(x-y) +i \delta k y}
  \nonumber \\ & & \hspace{1.5cm}
  + e^{ik(x-y)+i\delta kx}
  \big ] \Big \} +H.c.\; .
\label{vl} \end{eqnarray}

To deduce explicitly from Eq. (\ref{vl}) an interaction that can create
rotating states we consider the case that the laser intensity
is so weak that the effective Rabi frequency $|\Omega_0|^2
/(4 \Delta)$ is much smaller than the trap frequency $\omega
_\perp$. This results in the following consequences:
(i) If the atoms absorb and emit photons within a pair of
laser beams with the same frequency
(first square bracket in Eq. (\ref{vl}))
transitions between states of different energies $H_{c.m.}$
are off-resonant and can therefore be neglected;
(ii) If the atoms absorb a photon from one pair of laser beams
and emit it into the other pair with different frequency,
we have to take into
account the frequency difference $\Delta \omega$ between the
two photons with different frequencies. It produces an energy shift
so that transitions
between states having the energy difference $\pm \hbar \Delta
\omega$ are resonantly enhanced whereas other off-resonant transitions
are suppressed. Here we choose
\begin{equation}
\Delta \omega = \omega_\perp
\end{equation}
so that only transitions between neighboring states of the harmonic
oscillators are dominant.

Mathematically this can be incorporated by switching to the
interaction picture with respect to $H_{c.m.}$, so that the
annihilation operators $a_i$ are replaced by $a_i \exp
[-i \omega_\perp t]$ for $ i=x,y$. In the interaction picture,
by performing the rotating wave approximation with respect to
$\omega_\perp$,
the exponential $\exp [ ik (x-y)]$ has the form \cite{peter}
\begin{eqnarray}
e^{ik(x-y)} &\approx& e^{- \eta^2 /2} : J_0(2 \eta
    \sqrt{q^\dagger q}) :
  \label{j0} \\
e^{-i\Delta\omega t} e^{ik(x-y)} &\approx& i e^{- \eta^2 /2}
    :J_1(2\eta
    \sqrt{q^\dagger q}) : \; ,
\label{j1} \end{eqnarray}
where for notational convenience we have introduced the operator
$q:= (a_x-a_y)/\sqrt{2}$.
The normally ordered Bessel functions are defined as
\begin{equation}
  :J_n(\zeta \sqrt{q^\dagger q}):\quad := \left ( {1\over 2}
  \zeta q^\dagger
  \right )^n \sum_{m=0}^\infty \frac{(-\zeta^2 /4)^m}{m! (m+n)!}
  (q^\dagger)^m q^m \; .
\end{equation}
Since the initial state for the problem discussed here is the
condensate occupying the ground state of the trap, the
Lamb-Dicke parameter $\eta :=k R_\perp$ is generally larger than
one for the current realizable condensates \cite{experimente}
which usually have transverse sizes $R_\perp > 1 \mu$m.
As a result, Eqs. (\ref{j0}) and (\ref{j1}) imply that the terms
containing $\exp [\pm ik(x-y)]$ in $V_L$ of Eq. (\ref{vl}) are
exponentially suppressed for large $\eta$ and therefore can be
omitted for realistic cases. A physical interpretation of this
effect can be given by realizing that $\exp (i k x) =
\exp (i \eta (a_x + a_x^\dagger))$ acts as a displacement operator
on the trap states. Thus, for a large value of $\eta$, it would cause a
large displacement of the initial ground state of the trap.
But since the interaction is weak the energy required for this
displacement is not fully provided so that only the low-energy
part of the displaced ground state can be realized.

On the other hand, the Lamb-Dicke parameter $\delta \eta
:= \delta k R_\perp$ corresponding to the exponentials
$\exp [i \delta k x]$ and $\exp [i\delta k y]$ in Eq. (\ref{vl})
depends on the angle $\theta$ between the laser beams
through $\delta k = k \theta$ and can
therefore be adjusted over a wide range. Here we require $\theta$ to be
very small so that $\delta \eta \ll 1$ is valid. This allows us to
approximate the two exponentials by $1+i\delta k x$ and
$1+i\delta k y$, respectively. The small angle between the
laser beams lead to a simple dipole potential $V_L$
which linearly depends on the combination of
the creation and annihilation vortex operators of the trap.

In addition, in the interaction picture, the term
$\exp [-i \delta \omega t]
\{ 1+i\delta k x\} $ can be approximated by $i \delta k R_\perp a_x
^\dagger /\sqrt{2}$ under the rotating-wave approximation.
A similar expression is obtained for
$\exp [-i \delta \omega t] \{ 1+i\delta k y\} $ so that the
final expression for the complete Hamiltonian in the
interaction picture reads
\begin{equation}
H_{int} = \frac{\hbar |\Omega_0|^2}{4 \bar{\Delta}} \left \{
   4 + i \delta k R_\perp \left [ c_-^\dagger - c_-
   \right ] \right \} \; .
\label{hint} \end{equation}

The physical content of  Eq. (\ref{hint}) is very clear.
The second term in Eq. (\ref{hint}) exactly
excites atoms in the ground state $|0 \rangle$
into the rotating $c_-$ mode. Therefore the
initial cylindrical symmetry is
broken by the interaction Hamiltonian (\ref{hint})
and a rotating state of the center-of-mass motion
of the ground state
atoms in the trap is created with each atom gaining an angular
momentum $\hbar$
Thus, if we start from the ground state $|0\rangle$
of the trap, this term produces a vortex state with
angular dependence
$\exp (i \varphi)$ in the coordinate representation.
The term proportional to $c_-$ is the
corresponding Hermitian conjugate term and decreases the
angular momentum by $\hbar$.
By ignoring the unimportant constant
phase shift in Eq. (\ref{hint}), and by returning to the
Schr\"odinger picture, the time evolution operator $U(t)$
of the condensate has the simple form
\begin{equation}
  U(t) = \exp \left \{ \frac{t}{T_v} [c_-^\dagger \exp (i \omega
  _\perp t) - c_- \exp (-i \omega_\perp t)] \right \}\; ,
\end{equation}
where the time factor
\begin{equation}
T_v := \frac{4 \bar{\Delta}}{|\Omega_0|^2 \delta k R_\perp} \; .
\end{equation} characterises the time scale required for the
creation of rotating states. When the time $t$ is much smaller than
$T_v$, the ground state evolves into the state $U(t) |0 \rangle$
which approximately is a linear superposition
of the first excited vortex state and the initial
condensate occupying the ground state $|0 \rangle$ of the trap.
With increasing interaction time t, higher vortex states are created.
Finally we have a coherent state for the vortex operator $c_-$
for a long interaction time. This rotating coherent state is
a coherent superposition of vortex states circulating
in the same direction at different angular velocities.
From the coordinate representation of the vortex states,
$ \langle\vec{x}|(c_-^\dagger)^n | 0 \rangle = 2^{n/2} (x+iy)^n
R_\perp^{-(n+1)}
\exp [-(x^2+y^2)/(2R_\perp^2)]$,
we can deduce the probability density
for the coherent rotating state,
\begin{equation}
  \rho(t) = |\sqrt{N}
  \langle \vec{x}| U(t) |0 \rangle |^2 = \frac{N}{\pi R_\perp^2}
  \exp \big \{-\frac{(\vec{x}-\vec{x}_0(t))^2}{R_\perp^2} \big \}\; ,
\label{lsg} \end{equation}
where $\vec{x}_0(t) \equiv \langle \vec{x} \rangle =
(R_\perp t/T_v) [\cos(\omega_\perp t) \vec{e}_x -
\sin(\omega_\perp t) \vec{e}_y$ determines the center-of-mass
trajectory of the condensate. Eq. (\ref{lsg}) describes a condensate
starting at the center of the trap and rotating with increasing
amplitude thereby preserving its shape (see Fig. 2).

So far we have theoretically presented
a simple scheme to create rotating
states from a trapped ground-state Bose-Einstein condensate by a
light-induced dipole potential.
To analyze the experimental feasibility, we use the
experimental data of $^{23}$Na Bose-Einstein condensate created
by the MIT group \cite{sodium} as an example. The mass of the
$^{23}$Na atoms is $M= 3.8 \cdot 10^{-23}$ g and the wavelength
of the $^{23}$Na D-line transition is 589 nm so that the wave
number of the laser beams is given by $k=1.06\cdot 10^{5}$cm.
For the MIT trap
the trap frequency in the x-y-plane is $\omega_\perp
\approx 2\pi \cdot 320$Hz so that the transversal trap size parameter
is found to be $R_\perp \approx
1.17 \mu$m.  With these parameters the Lamb-Dicke parameter
$\eta = k R_\perp$ is approximately 12.4 so that the right hand
side of Eqs. (\ref{j0},\ref{j1}) can safely be ignored
for low lying trap states. Hence the $^{23}$Na Bose-Einstein
condensate is an appropriate candidate to satisfy the conditions
required above.

The small Lamb-Dicke
parameter $\delta \eta = k \theta R_\perp$ is determined
by the angle $\theta$ and we can realize a value $\delta \eta=0.1$ by
adjusting $\theta$ around 0.46 degrees. In addition,
to make sure that only resonant
interaction is included, the effective Rabi frequency
$\Omega_{eff} := |\Omega_0|^2/(4\bar{\Delta})$ must be much smaller
than the trap frequency $\omega_\perp$. This can be experimentally
realized if the laser intensity is chosen so weak that
$\Omega_{eff}$ is about 200 Hz. In this case, the characteristic
time scale for vortex excitation
$T_v$ is about 50 ms which is experimentally reasonable
since it is much shorter than the life-time of the
currently realized trapped condensates.

In conclusion we have shown that it is possible to create
rotating states from a trapped ground-state Bose-Einstein condensate
in  a trap by using light-induced forces. Technically we
show that dipole potential induced by
four travelling-wave laser beams with an appropriate configuration
in space, phase and frequency can be used to realize such
a purpose. The important features of the proposed scheme
are the following: (i) The low intensity of the
laser beams  to guarantee that only resonant transitions can
occur; (ii) The small frequency
difference between the two pairs of lasers to provides the
energy difference required to bring the atoms from the ground state
of the trap into a superposition of the first excited states.
(iii) The small angle $\theta$ between the laser beams
avoids the problem that the Lamb-Dicke parameter $\eta$
is so large that the transitions are strongly suppressed
if the difference between the absorbed and re-emitted
momentum is not very small.
(iv) The phase difference between the laser beams leads to a
special form of the light-induced potential $V_L$ which
exactly results in a creation of vortex states.

Although the nonlinear interatomic interaction between the
atoms is ignored in our discussion, it can be shown that the total
Hamiltonian including the nonlinear interatomic interaction commutes
with the second quantized form of the angular momentum operator
$J_z$. Hence in principle the interatomic
interaction does not affect the creation of rotating states, and only
affects the spatial profile of the macroscopic wave function in the
radical direction. The detailed discussions on the effect
of the nonlinear interatomic interaction will be
covered in a long publication \cite{peter}.
%%%%%%%%%%%%%%%%%%%%%%%%%%%%%%%%%%%%%%%%%%%%%%%%%%%%%%%%%%%%%%%%

{\bf Acknowledgement}: The work has been supported by Australian
Research Council. K.-P. M. wishes to thank J\"urgen Audretsch
and his theory group, where this work was started,
for many thought provoking conversation and the
collegial atmosphere during the last few years.
W. Z. thank E. Wright for helpful discussions.
%%%%%%%%%%%%%%%%%%%%%%%%%%%%%%%%%%%%%%%%%%%%%%%%%%%%%%%%%%%%%%%%
%\newpage

\newpage
{\bf Figure Captions:}\\[1cm]

{\bf Fig. 1:} Proposed scheme to create a rotating state being a
coherent superposition of vortex states of the condensate.
The laser beams 1 and 2 are oriented along the x- and y-axis.
A frequency modulator produces a slight frequency difference between the
beams (1,2) (solid line) and (3,4) (dashed line).
A small angle between beam 1 and 3, and beam 2 and 4 is arranged by
the adjustable mirror.
\\[2cm]
{\bf Fig. 2:} The trajectory of the center-of-mass of the condensate in the
x-y-plane under the influence of the four laser beams described above.
x and y are given in units of the trap size $R_\perp$.
During the rotation the shape of the condensate remains a Gaussian profile
with a width of $2R_\perp$.
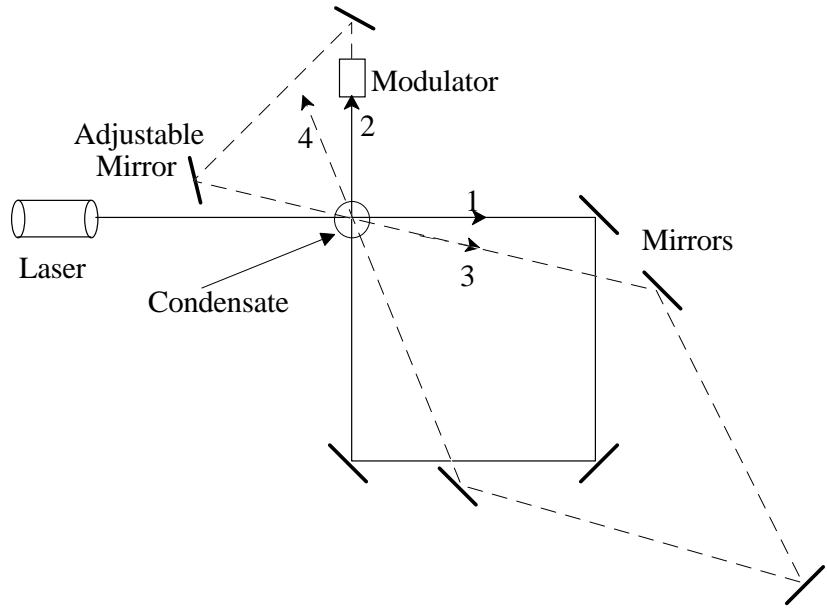
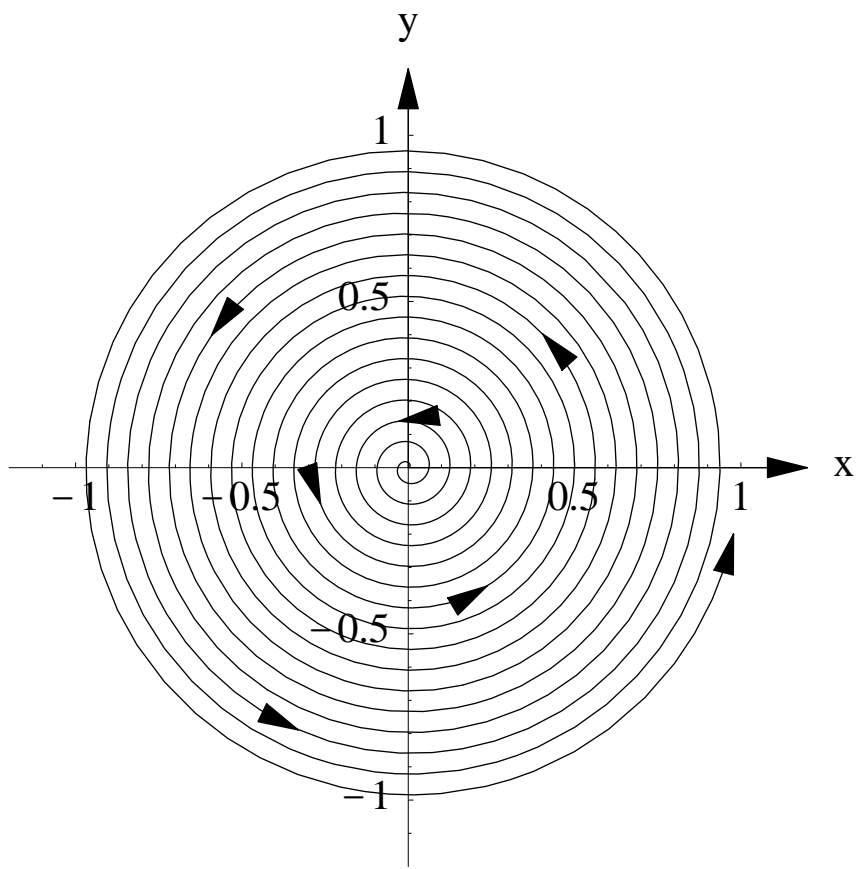
\end{document}